\documentclass[conference]{IEEEtran}
\IEEEoverridecommandlockouts
\usepackage{cite}
\usepackage{amsmath,amssymb,amsfonts}
\usepackage{algorithmic}
\usepackage{comment}
\usepackage{graphicx}
\usepackage{textcomp}
\usepackage{mathtools}
\usepackage{float}

\def\BibTeX{{\rm B\kern-.05em{\sc i\kern-.025em b}\kern-.08em
    T\kern-.1667em\lower.7ex\hbox{E}\kern-.125emX}}
\begin{document}

\title{Towards a Lightweight Continuous Authentication Protocol for Device-to-Device Communication}
\author{\IEEEauthorblockN{Syed W. Shah, Naeem F. Syed, Arash Shaghaghi, Adnan Anwar, Zubair Baig, and Robin Doss}
\IEEEauthorblockA{\textit{Deakin University, Geelong, Australia} \\
\textit{Centre for Cyber Security Research and Innovation (CSRI)}\\
\{syed.shah, naeem.syed, a.shaghaghi, adnan.anwar, zubair.baig, robin.doss\}@deakin.edu.au \\}
}

\maketitle

\begin{abstract}
Continuous Authentication (CA) has been proposed as a potential solution to c	ounter complex cybersecurity attacks that exploit conventional static authentication mechanisms that authenticate users only at an ingress point. However, widely researched human user characteristics-based CA mechanisms cannot be extended to continuously authenticate Internet of Things (IoT) devices. The challenges are exacerbated with increased adoption of device-to-device (d2d) communication in critical infrastructures. Existing d2d authentication protocols proposed in the literature are either prone to subversion or are computationally infeasible to be deployed on constrained IoT devices. In view of these challenges, we propose a novel, lightweight and secure CA protocol that leverages communication channel properties and a tunable mathematical function to generate dynamically changing session keys. Our preliminary informal protocol analysis suggests that the proposed protocol is resistant to known attack vectors and thus has strong potential for deployment in securing critical and resource-constrained d2d communication.
\end{abstract}

\begin{IEEEkeywords}
Zero Trust Architecture (ZTA), Continuous Authentication, Device-to-Device, IoT
\end{IEEEkeywords}

\section{Introduction}\label{intro}
Traditional perimeter--based security architectures do not have sufficient design traits to foster security for critical infrastructures such as smart girds, industrial IoT (IIoT) and other cyber physical systems (CPS). In view of this, Zero Trust Architecture (ZTA), which is based upon the notion of least privilege is widely seen as an appropriate alternative \cite{rose2019zero}. As referred to in the National Institute of Standards and Technology (NIST) report on ZTA, authentication and access control are two key tenets of ZTA for all CPS infrastructures. For a reliable implementation of ZTA, authentication is considered to be of utmost importance and a fundamental building block. Traditional authentication mechanisms only determine a user's identity at login time. In the ZTA context, however, there is no implicit trust that exists between the end points. Consequently, a Continuous Authentication (CA) scheme that aims to mutually authenticate end points continuously during the communication session would play a crucial role to enable security.  Most contemporary CA schemes are suitable only for human/user authentication as they rely on a users' behavioural biometric traits such as typing or key-tapping behaviors \cite{TFIS2013, DSS2014, CS2015}. \par

The rapid growth and adoption of Internet of Things (IoT) and edge computing platforms have led to a \emph{Device-to-Device (d2d)} authentication requirement. Furthermore, the critical nature of various operational technology (OT) infrastructures would require a continuous confirmation of identities of communicating devices. This is because, in critical infrastructures, malicious data that penetrates a network beyond the entry-point authentication mechanism in place, may pose serious risks. Despite the need for continuous d2d authentication, very few solutions have been proposed. One of the reasons for this may be the resource constrained nature of IoT devices in terms of computing power and memory. Extant schemes that deal with d2d authentication have incorporated device fingerprints that require additional hardware \cite{chen2016s2m} or complex initial processing \cite{yu2019continuous} or those that operate in unstable environments \cite{chuang2018lightweight}. In addition, IoT edge authentication protocols that do not use any device fingerprints and only rely on time-bound key generation techniques \cite{sathyadevan2019protean} have also been proposed.

In this paper, we propose a novel continuous authentication protocol suitable to foster device to device communication in resource-constrained devices. Our proposed solution operates in two phases: a \emph{mutual-authentication} phase and a \emph{continuous-authentication} phase. During the mutual authentication phase, devices agree on a key that is based upon contextual information of devices. More precisely, the contextual information required for defining the secret key is derived from wireless channel characteristics of the communicating devices (i.e., channel state information), dynamically refreshed per session. Both devices securely exchange a portion of a shared secret, for continuous authentication in such a way that the secret key cannot be reverse engineered. To enable practical implementation on resource-constrained devices, we employ simple computational operations such as exclusive OR \emph{(XOR)} and Hash-based Message Authentication Codes (HMAC). \par

The rest of this paper is organised as follows.  \S\ref{relatedwork}, presents a succinct review of related work, while \S\ref{protocol} provides an overview of the proposed protocol and underlying assumptions.  \S\ref{detailsoftheprotocol} contains the details of our proposed continuous authentication protocol. We informally analyse the security of our proposed protocol with respect to the six most common attacks reported in the related literature in \S\ref{informal}, make concluding remarks in \S\ref{conclusion}.

\section{Related Work} \label{relatedwork}
In \cite{sathyadevan2019protean},  a lightweight authentication scheme that uses time-bound authentication keys for constrained devices deployed in IoT networks, is proposed. Secret keys are refreshed at the end of each session, and are synchronized both at an edge-node and a gateway. The evaluation of this approach confirms it is lightweight and robust against most adversarial threats. However, this approach does not address the much needed continuous d2d authentication which is an essential tenet for ZTA. \par 

Another interesting approach for accomplishing d2d authentication that is currently gaining momentum is referred to as \emph{device fingerprinting}. Hardware characteristics uniquely identify a particular device in such schemes. For example, the authors in \cite{chen2016s2m} proposed a lightweight device authentication protocol that leverages the frequency response of acoustic hardware (speaker/microphone) as a fingerprint. They demonstrated that an authenticating device can generate an audio signal and transmit it through its speakers to the receiver - which has already fingerprinted the source device during an initial enrollment phase and saved its fingerprint. In the authentication phase, the verifier processes the received sound and computes a frequency response to see whether the sound is being transmitted by the same device which was previously enrolled during the learning phase. This approach has demonstrated good accuracy and robustness against cyber attacks. However, this scheme is only tested on mobile phones and its efficacy on resource-constrained devices is not known. In addition, it also has a limited operating range of 5m, and may not be suitable for d2d authentication in situations where devices are located far apart. Furthermore, its appropriateness for accomplishing continuous authentication is also not analyzed. \par

Authors in \cite{chuang2018lightweight} proposed a lightweight continuous d2d authentication scheme that utilizes the battery capacity values of the sensor node for continuous authentication on a corresponding gateway. During the enrollment phase the sensing device can exchange some parameters with the gateway, based on which the gateway can estimate the time-dependent values of a sensor's battery capacity. Whenever a sensor-node wants to send data to the gateway, it will also send a value presenting the current battery capacity, which is checked by the gateway against its own estimate. If two values exhibit a reasonable match, then the device is successfully authenticated. The proposed approach makes an unrealistic assumption - i.e., sensors nodes are always battery powered. Moreover, the authors modelled the battery estimate using a linear function. This implies that an adversary would essentially require only two values of this function to be able to correctly guess all other possible values. \par

The protocol suggested in \cite{yu2019continuous} relies on fingerprinting a device using Channel State Information (CSI). The authors demonstrated that the Carrier Frequency Offset (CFO) obtained from a CSI can uniquely fingerprint any device and thus may be used for continuous d2d authentication. In the initialisation phase, communicating devices need to exchange multiple data frames between themselves to accurately estimate the fingerprint, and then a fingerprint extraction algorithm is used to extract the tuple of CFO presented as unique features. In authentication phase, a purpose-built algorithm is deployed for matching the received fingerprint with the saved one. The limitation of this work lies with the extensive processing pipeline requirement imposed, making it difficult for adoption in resource-constrained environments.

\section{Protocol Fundamentals and Notations} \label{protocol}

\subsection{Assumptions}
The proposed protocol has been defined under the following assumptions:
\begin{enumerate}
    \item The initialisation phase is assumed to be completed in a secure environment with the secret key and identifiers exchanged during this process considered to be secure.
	\item Edge devices are assumed to be resource constrained in terms of storage and processing capabilities.
    \item D2D communication can occur between two resource-constrained devices or between a resource-constrained device and a resourceful device such as a gateway. For simplicity, the proposed protocol is explained for an edge-to-gateway scenario. However, the protocol is designed in such a way that it may easily be customised for d2d scenarios as well.
\end{enumerate}

\subsection{Protocol Phases}
The protocol comprises three phases, described as follows:
\begin{enumerate}
    \item \emph{Initialisation Phase:} In this phase both the communicating devices retrieve the identities of each other and generate authentication information that enables the subsequent phases.
	\item \emph{Mutual Authentication Phase:} In this phase, both devices mutually authenticate each other and agree on a session key derived through processing of the Channel State Information (CSI).
    \item \emph{Continuous Authentication Phase:} In this phase, both communication devices leverage CSI-based keys that were previously derived, to agree on a shared secret and subsequently incorporate the key into the algorithm, to enable it to perform continuous mutual authentication.
\end{enumerate}

\subsection{Notations}
The notations applied for the protocol are presented in Table \ref{notations}.

\begin{table*}[!t]
    \centering
  \caption{Notations used in defining the protocol}
    \resizebox{\textwidth}{!}
    {
    \begin{tabular}{p{2cm}|p{13cm}}
         Symbol & Description \\
        \hline 
         $ID_{edge}$ & Edge device ID \\
         $ID_{gw}$ & Gateway device ID \\
         $Seed$ & Initial Seed to generate random numbers using PRNG (pseudo-random number generator) \\
         $E_{init}$ & Encryption Key exchanged during initialisation or device enrollment \\
         $E_{Key}$ & Encryption Key \\
         $D_{Key}$ & Decryption Key \\
         $r$ & a random number on both edge-node and gateway generated through a common seed \\
         $C_{r}$ & Channel State Information measured from the packet sent by the gateway \\
         $C_{i}$ & Channel State Information measured from the packet sent by the edge \\
         $HMAC_{j} (.)$ & Hash Based message authentication code generated using a secret key $j$ \\
         $\bigoplus$ & Bitwise exclusive-OR operator \\
         $H(.)$ & One-way hash function \\
         $T$  & Session duration in seconds set between edge and gateway \\
         $a$ and $b$ & Random number generated during continuous authentication phase within certain bounds \\
         $M_{i}$   & $i_{th}$ intermediate messages \\
         $t_{m}$ & Time stamp at end of mutual-authentication phase \\
         $t_{c}$ & Time stamp within continuous authentication phase \\
         $Ctr_{e}$ & Counter Value on edge-node \\
         $Ctr_{g}$ & Counter Value on gateway \\
         $f$ & Linear or a non-linear function used in continuous authentication phase \\
         $SN_{key}$ & Encryption/Decryption key used in continuous authentication phase of a session \\
    \end{tabular}}
  
    \label{notations}
\end{table*}

\section{Continuous Authentication Protocol} \label{detailsoftheprotocol}

In this section, we present a detailed description of all three phases of our proposed protocol - i.e., initialization, mutual authentication, and continuous authentication. \par

\subsection{Initialisation Phase}
Edge devices that wish to communicate with the gateway are first enrolled during the initialisation phase. In this phase, parameters that are required for mutual authentication are setup on both the gateway and the edge device. To begin with, the identifier label of an edge device ($ID_{edge}$) is exchanged with the gateway. Similarly, the identifier of the gateway node ($ID_{gw}$) is exchanged with the edge device, both occurring over a secure channel.\par

After the exchange of identifiers, a random number ($r$) is generated using a Pseudo-Random Number Generator (PRNG) on both the edge and gateway, derived from a common $Seed$ value. Based on the parameters $ID_{edge}$ and $r$, a common secret key $E_{init}$ is generated by both edge and the gateway, which is subsequently used by the two devices to initiate the mutual authentication phase. At the end of the device enrollment phase, the edge device stores the $E_{init}$, $ID_{gw}$ and the $seed$ values. Similarly, the gateway stores the $E_{init}$, $ID_{edge}$ and $Seed$ values in a database. Since the gateway is assumed to interact with multiple edge devices and has stronger capabilities in terms of resources, it is entrusted to store secure keys and random numbers for each enrolled device, which can be retrieved for the subsequent stages of authentication. It is assumed that the device enrollment process is performed under supervision in a controlled and secure environment to protect the integrity of parameters exchanged as indicated in \cite{sathyadevan2019protean}. \par

\subsection{Mutual-Authentication Phase}

In this phase, devices authenticate each other as a first step prior to initiating data communication. To initiate mutual authentication, security parameters are exchanged during the device enrollment step. When an edge device has data to transmit, it commences the authentication process with the gateway device. In order to successfully authenticate, the edge device sends the hash of its ID ($ID_{edge}$) to the gateway, which processes the same to retrieve the corresponding security key ($E_{init}$) and random number ($r$) from its database. 
The gateway then computes the CSI for the packet received from the edge, $C_{i}$. Once the gateway verifies the edge identifier tag and is able to match the $ID_{edge}$ with a database entry, it sends a response packet containing its identifier ($ID_{gw}$) to the edge device. The edge device then extracts the CSI value from the packet received from the gateway, $C_{r}$. Since the CSI values $C_{r}$ and $C_{i}$ will be used later to generate security keys, the edge device computes $m_{a} = C_{r} \oplus H(E_{init} \oplus r)$, which prevents it from sending $C_{r}$ as clear text.  If $m_{a}$ is intercepted, the $C_{r}$ value cannot be directly computed without the knowledge of both the random number $r$ as well as the secure key $E_{init}$. To further prevent tampering of the the edge's reply to the gateway, an $HMAC$ of the message is also sent, $M3$. Based on this response from the edge, the gateway verifies the identity of the edge device by computing $M_{3}'$ using the secure key $E_{init}$ and random number $r$, which were generated from a common seed value (shared between the two devices). If the values of $M_{3}'$ and $M_{3}$ are found to be the same, this indicates that the sender is a legitimate edge device and that the messages have not been modified by an adversary, thereby succeeding in authenticating the edge device. If the messages $M_{3}'$ and $M_{3}$ do not match, then the gateway terminates the protocol, as a hash mismatch points towards a message tampering attack or a communication line fault.

Once the edge device is authenticated, the gateway extracts the $C_{r}$ value by first computing $m_{a} \oplus H(E_{init} \oplus r)$. The gateway then sets the secure key as $C_r$ and computes $m_b$ as $C_i \oplus H(E_{key} \oplus r)$. The gateway then sends $M_4$, which is computed as $HMAC_{E_{key}}(m_b,H(ID_{gw},r))$ and sends the values $m_b$, $M_4$ and $H(ID_{gw})$ to the edge device. At this point, the edge also sets the secure key as $C_r$ and computes the value of $M_4'$. If $M_4$ and $M_4'$  match, then this verifies the identity of the gateway and ensures that the message sent by the gateway is not modified during transmission. Once the gateway is authenticated, the edge sets the acknowledgement ($Ack$) value to $1$. 

The edge further extracts the value of $C_{i}'$ from $m_{b}$. Using the obtained $C_r$ and $C_i$ values, the edge sets the new secure key as $SN_{key}$ as $C_r \oplus C_{i}'$. For future mutual authentication phases, the $E_{init}$ is updated to use $SN_{key}$ to ensure freshness of keys for each step of the mutual authentication phase.xt In addition, a new seed value is generated to prevent the use of compromised random numbers in latter mutual authentication phases.
Counters $Ctr_e$ and $Ctr_g$ are adopted for the mutual authentication phase to avoid replay attacks and to ascertain freshness of the packets.
The updated $Seed$, $ACK$, and $C_{i}'$ values are encrypted using the secure session key $SN_{key}$, subsequently transmitted to the gateway, $M_5$. The mutual authentication phase is considered successful when the gateway receives the $ACK$ value of $one$ from the legitimate edge device, and $C_{i}'$ matches $C_i$. The gateway then sets the session duration \(T\) and timestamp value $t_m$ to record the time when mutual authentication was noted as successful. Else the authentication is aborted by the gateway.

    \begin{figure*}[]
    \centering
 
            \begin{tabular} {| l | l |}
                \hline
                \centering
                \hspace{1.5cm}\textbf{Edge [$E_{init},ID_{edge}, ID_{gw},Seed\;$]}  &  \hspace{1.5cm}\textbf{Gateway [$E_{init},ID_{gw}, ID_{edge}, Seed\;$]} \\
                \hline
                Send the  $ID_{edge}$   & \\
                \multicolumn{1}{|c|}{\hspace{4.5cm}$H(ID_{edge}) $} & \\
                \multicolumn{1}{|r|}{$------>$} &  Retrieves $E_{init}$ and $Seed$ from database using the $H(ID_{edge})$ \\
                 
                 & if $ID_{edge}$ is in the database \\
                 & \hspace{0.5cm}  Computes the CSI of the received packet $\rightarrow$ $C_i$ \\
                 & \hspace{0.5cm}  $r\; \leftarrow\; PRNG(Seed)$\\
                 & \hspace{0.5cm}  Send response packet with $ID_{gw}$\\
                 & else \\
                 & \hspace{0.5cm} Go to device initialisation phase\\
                   
                   & \hspace{0.6cm}$H(ID_{gw})$ \\
                   & $<------$\\
                   if $ID_{gw}$ matches the stored gateway ID: & \\
                   
                \hspace{0.8cm} Computes the CSI of the received packet $\rightarrow  \; C_r$ & \\
                \hspace{0.8cm} $r\; \leftarrow\; PRNG(Seed)$&\\
                \hspace{0.8cm} $m_a \; \leftarrow \;C_r \; \oplus \; H(E_{init} \; \oplus \; r)$  & \\
                \hspace{0.8cm} $M_3\; \leftarrow\; HMAC_{E_{init}} (H(ID_{edge}),m_a,r)$ & \\
               else & \\
                \hspace{0.8cm} $Abort$ & \\
                \multicolumn{1}{|r|}{$H(ID_{edge}), m_a, M_3 $} & \\
                \multicolumn{1}{|r|}{$------>$} & 
                
                 Compute $M_3'\; \leftarrow \;HMAC_{E_{init}}(H(ID_{edge}),m_a,r)$ \\
                & if ($M_3' == M_3$ ) \\
                & \hspace{0.5cm} Compute $C_r' \;\leftarrow \;m_a  \; \oplus H(E_{init} \; \oplus \; r)$ \\
                & \hspace{0.5cm} $E_{key} \; \leftarrow \; C_r'$ \\
                & \hspace{0.5cm} $m_b \leftarrow\; C_i \; \oplus \; H(E_{key}\; \oplus \; r)$ \\
                & \hspace{0.5cm} $M_4 \;\leftarrow \;HMAC_{E_{key}}(m_b,H(ID_{gw}),r)$ \\
                & else \\
                & \hspace{0.5cm}$Abort$ \\
               
                & \hspace{0.1cm} $H(ID_{gw}),M_4,m_b$ \\
                & $<------$ \\
                Set $E_{key}\; \leftarrow \;C_r$ &\\
                Compute $M_4' \leftarrow \; HMAC_{E_{key}}(H(ID_{gw}),m_b,r)$ & \\
                if ($M_4'$ == $M_4$) & \\
                \hspace{0.5cm} $Set\; ACK(1)$ & \\
                \hspace{0.5cm} Compute: $C_i'\; \leftarrow\; m_b \; \oplus \; H(E_{key}\;\oplus\;r) $ & \\
                \hspace{0.5cm} $SN_{key}\;\leftarrow\;C_r\oplus\;C_i'$ & \\
                \hspace{0.5cm} $E_{init}\; \leftarrow\; SN_{key}$ and $Seed\; \leftarrow\; rand(int)$ & \\
                \hspace{0.5cm} Set $Ctr_e\;\leftarrow\;1$ &\\
                \hspace{0.5cm}    $M_5\;\leftarrow\;SN_{key}(C_i', Seed, ACK)$ & \\
                else & \\
                \hspace{0.5cm} $ Abort$ & \\
                
                \multicolumn{1}{|c|}{\hspace{4.8cm}$M_5$} & \\
                \multicolumn{1}{|r|}{$------>$} & 
                 $SN_{key}\; \leftarrow\; C_r' \oplus C_i$ \\
                & Decrypt $C_i'\; Seed'\; ACK = SN_{key}(M_5)$ \\
                & if ($ACK == 1\; \&\& \;C_i' == C_i$)\\
                & \hspace{0.5cm} Set $Seed\; \leftarrow\;  Seed'$ and $Ctr_g \; \leftarrow \; 1$\\
                & \hspace{0.5cm} $E_{init}\; \leftarrow\; SN_{key}$\\
                & \hspace{0.5cm} Set session duration $T$ and timestamp $t_m$  \\
                & \hspace{0.5cm} \textbf{Authentication successful} \\
                & else \\
                & \hspace{0.5cm} $Abort$ -  \textbf{Authentication unsuccessful} \\
               
                \hline
                \hline
                
            \end{tabular}
        \caption {Proposed Protocol for Mutual-Authentication Phase }
        \label{Fig_mut}
    \end {figure*}

\subsection{Continuous Authentication Phase}
 Continuity, in general, refers to establishment of a secure communication session for a certain duration of time (i.e., T in our case). In principle, it supplements the mutual authentication phase to make sure that the device that was authenticated at the inception of a session remains the same throughout, and this process is referred to as Continuous Authentication. \par
 
 To strike the balance between security and computational cost during the continuous authentication phase, we propose a new approach that makes use of the device context information (i.e., CSI for generating the key ($SN_{key}$)) along with a shared secret which dynamically varies between a linear or a non-linear function depending upon the partially set exponent values, and by securely exchanging the same between the two devices at the start of the continuous authentication phase. This provides two main benefits. First, the exponent values of the function are set in such a way that these values are easy to compute on constrained devices by restricting exponents within certain bounds depending upon the capabilities of devices. Second, the exponent values change dynamically for each new session (i.e., after T secs). This ensures that the function is not compromised and its values are aligned to the resource constraints of the device. In addition, during the continuous authentication phase, exponents and function values are exchanged securely between the devices using the dynamic session key (i.e., $SN_{key}$), which is based on the context information of each device (i.e., CSI), and is thus unlikely to be compromised. Furthermore, we make use of time stamps along with local counters on both devices to make the protocol more secure during this phase. Note that, although we use time stamps for achieving security properties, the protocol itself does not necessitate a precise time synchronization step between the devices. \par
 
 Figure \ref{Fig_cont} depicts the proposed protocol for continuous authentication. At the start of the continuous phase, the edge-device randomly selects an exponent value (i.e., \emph{\lq a\rq}). It then computes $m_c$ by computing the \emph{xor} of this value and the hash of ($SN_{key} \oplus r$), and sends this along with ($ID_{edge}$) (i.e., $M_6$) securely to the gateway by first encrypting the same with a session key ($SN_{key}$). The gateway upon receiving ($M_6$) decrypts it using $SN_{key}$ and obtains $ID_{edge}$ and $m_c$, to enable computation of the exponent \emph{\lq a\rq}. It then sets the current time stamp to $t_c$ and computes its difference from the time stamp $t_m$, which was set during the mutual authentication phase. If the difference is greater than the session duration \emph{T}, the gateway sets \emph{ACK} value to $0$ so as to trigger the mutual authentication phase. If the difference between the time-stamps is within the allowed time $T$, the gateway checks whether $ID_{edge}$ is found within its database. If the edge node is found, the gateway then proceeds to set the second exponent (i.e., \emph{\lq b\rq}) randomly. Next, the gateway computes $m_d$ by computing the $xor$ between \emph{\lq b\rq} and hash of ($SN_{key} \oplus r$), $f = t^a + t^b$, and increments the value of the local counter. It then sends a message $M_7$ back to the edge-node by encrypting $ACK, f, t, a, m_d, Ctr_g$ with ($SN_{key}$). \par 
 
 The edge-node decrypts $M_7$ and verifies the $ACK$ value, and goes back to mutual-authentication, if its value is $0$. If not, the edge-node checks whether the exponent value \emph{\lq a\rq} sent back by the gateway is the same as the one that it had previously sent. This ensures that the message $M_7$ comes from the device with which the edge-node has exchanged the exponent value \emph{\lq a\rq} by leveraging the dynamic context-dependent session key (i.e., $SN_{key}$). It also checks the values of the local counter as well as the value sent by the gateway so as to make sure that messages are being received in the correct order. Then, the edge-node obtains the exponent \emph{\lq b\rq} by retrieving $m_d$ from the received message $M_7$, and computes the value of $f$ locally (i.e., $f'$) by retrieving values of \emph{\lq a\rq} and \emph{\lq b\rq}, to compare the same against the value received from the gateway. A positive match of these values ensures that the edge-node is communicating with the same device with which it securely exchanged the exponent \emph{\lq a\rq}. The edge-node then responds by sending $M_8$ that contains $f'$ and an incremented value of the local counter back to the gateway, encrypted with session key $SN_{key}$. The gateway upon receiving $M_8$ decrypts it, and verifies whether the value of $f'$ is the same as that of $f$, and also matches the counter values. If these parameters are verified, the gateway goes back \emph{Point X} to update the time stamp $t_c$ and again compares it with $t_m$, to verify whether the difference between the time stamp values lies within the session time $T$. If the differences are found to be within the allowed time range \emph{T}, then the same procedure is repeated to accomplish the continuous authentication.  
 
    \begin{figure*}[]
    \centering
   
           \begin{tabular} {| l | l |}
                \hline
                \centering
                \hspace{1.5cm}\textbf{Edge [$SN_{key},r,Ctr_{e}$]}  & \hspace{1.5cm}\textbf{Gateway [$SN_{key},t_m,r,Ctr_{g}$]} \\
                \hline
               
               Set $a = rand(init)$ & \\
               $m_c \leftarrow a \oplus H(SN_{key} \oplus r)$&\\
               $M_6  \leftarrow {SN_{key}(m_c,ID_{edge})}$ & \\
               \hspace{4.5cm} $M_6$  & \\
                 \hspace{3.7cm}$------->$& 
                 $m_c, ID_{edge} \leftarrow SN_{key}(M_6)$ \\
                & $a' \leftarrow{ m_c \oplus H(SN_{key} \oplus r)}$\\
                & \textbf{Point X} \\
                & Set timestamp  $t_c$ \\
                & if $(t_c \; -\; t_m) \; > \; T$ \\
                & \hspace{0.5cm}Set $ACK,\; f,\; t,\; m_d \leftarrow 0$  \\
                & \hspace{0.5cm}\emph{Send $M_7$ and Go to Mutual Authentication Phase} \\
                & $elseif \, ID_{edge}$ in database: \\
                & \hspace{0.5cm}Set $b = rand(init)$ \\
                & \hspace{0.5cm}$Ctr_{g} ++$\\
                & \hspace{0.5cm}Set $t \leftarrow\; t_c - t_m$  \\
                & \hspace{0.5cm}Compute: $f \leftarrow t^a + t^b$  \\
                & \hspace{0.5cm}$m_d \leftarrow {b \oplus H(SN_{key} \oplus r)}$\\
              & \hspace{0.5cm}Set $ACK \leftarrow\; 1$  \\
                & $else$ \\
                & \hspace{0.5cm} $Abort$ \\
                & $M_7 \leftarrow{SN_{key}(ACK,f,a',t,m_d,Ctr_g)} $\\ \\
                &\hspace{0.8cm} $M_7$\\
                &$<-------$\\
                
                $ACK,\;f,\;a',\;t,\;m_d,\;Ctr_g$ = $SN_{key}(M_7)$&\\
                if ($ACK == 0$) &\\
                \hspace{0.5cm}\emph{Go back to Mutual Authentication Phase} & \\
                $elseif \, (a' == a)\, \&\& \, (Ctr_e + 1 == Ctr_g)$ &\\
            
                \hspace{0.5cm} $Ctr_{e} ++$ &\\
                \hspace{0.5cm} $d \leftarrow{ m_d \oplus H(SN_{key} \oplus r)}$ & \\
                \hspace{0.5cm} $b \leftarrow d$ &\\
                \hspace{0.5cm} Compute: $f' \leftarrow\; {t^a + t^b}$ & \\
                \hspace{0.8cm}$if (f' == f) $ &\\
                  \hspace{0.95cm}$M_{8} \leftarrow{SN_{key}(f',Ctr_e)}$ &\\ &\\
                \hspace{0.8cm}$else$ &\\
                  \hspace{0.95cm}$Abort$ &\\
                 
                \hspace{4.5cm} $M_8$  & \\
                \hspace{3.7cm}$------->$& 
                 $f'' , Ctr_e = SN_{key}(M_{8})$ \\
              &   $if (f'' == f)\, \&\& \,(Ctr_g == Ctr_e) $ \\
              & \hspace{0.5cm}$Go\; to\; Point\; X$\\
              &   \hspace{0.50cm} $Abort$  \\

                \hline
                 \hline
            \end{tabular}
        
        \caption {Proposed Protocol for Continuous Authentication Phase }
        \label{Fig_cont}
    \end {figure*}

\section{Informal Security Analysis} \label{informal}
In this section, we provide an informal analysis of the security properties of the proposed continuous authentication protocol. 

\subsection{Replay Attack}
In a replay attack, an adversary eavesdrops transmitted messages, and later replays some of these to impersonate a legitimate device. During the mutual authentication phase, a previously recorded message when replayed will not violate the security property of the scheme, as each time a new random number and CSI value are generated and adopted. For example, if an attacker replays a previously recorded $H(ID_{edge})$ message to the gateway, and receives a response from the same, both devices will record the CSI. However, the edge-node will not be able to craft $m_a$ and $M_3$ such that they are verified on the gateway, as both these values rely on freshly generated random numbers which are known only to valid devices and also on a previously saved session key, $E_{init}$. Likewise, if a compromised edge-node sends a previously recorded message (i.e. $m_a, M_3$) to a gateway, it will not be verified as being legitimate, as the gateway will compute $M_3'$ with a fresh random number and fail to match with $M_3$. Even if this check is somehow compromised, and the gateway responds to it by sending $H(ID_{edge})$, $m_b$ (i.e., $C_i \oplus H(E_{key}\oplus r)$ back to the malicious node; where $E_{key} = C_r$, and $M_4$ (i.e, $HMAC_{Ekey}(H(ID_{edge}),m_b,M_4)$), the adversary will not be able to compute $C_i$ from the received message as it would require knowledge of $C_r$ from previously recorded messages as well as a fresh random number resulting in a failure to complete the mutual authentication phase. Likewise, a previously recorded $M_5$ message cannot be replayed as it will fail decryption on the gateway which uses freshly generated CSI based $SN_{key}$ to decrypt it. In the CA phase, time-dependent linear or non-linear function values alongside the counter values ensures resistance against the replay attacks. A previously recorded message, $M_6$, or $M_7$ cannot not be decrypted on the other device as fresh value of $SN_{key}$ is used. Messages $M_7, M_8$ replayed within a valid session (time < T) will not have  matching counter values, therefore the replay attack will not succeed.

\subsection{Impersonation Attack}
During an impersonation attack, a malicious entity may attempt to masquerade as a legitimate edge node (or gateway). In the mutual authentication phase, an impersonating device will not know the random number ($r$) that is used for computing $C_i$ and $C_r$ values locally, merely based on the message sent by the other communicating party, and therefore will be unable to generate the key $SN_{key}$, consequently causing the attack to fail. In the proposed CA phase, the impersonating device would need to know the exponents of the linear or non-linear function, for which knowledge of the session key ($SN_{key}$) is essential. As an impersonator is unable to gain knowledge of dynamically changing keys for each session that depend upon freshly measured CSI value, any such attacks will be thwarted. In addition, the key only works for a set duration of \emph{T} secs, and exponents of the linear or non-linear functions change after a fixed duration. Hence having compromised keys or exponent values will not threaten the security property of the scheme for the full device lifetime.

\subsection{Man-in-The-Middle Attack} 
In a Man-in-The-Middle (MiTM) attack, an active attacker furtively relays and manipulates the messages exchanged between device pairs thereby convincing them to believe that they are directly communicating with each other. During the scheme's mutual authentication phase, an MiTM attacker would require knowledge of both $C_i$ and $C_r$ values along with knowledge of the recently generated random number of the legitimate communicating devices. As knowing all this is essential but difficult for an active attacker that has placed itself on the communication line between the legitimate communicating devices, the protocol will be secure against MiTM attacks during this phase. For example, if an MiTM attacker attempts to manipulate the message ($H(ID_{edge}, m_a, M_3)$) and relays it to the gateway, it will not be verified as being genuine on the gateway, as the locally computed value of $M_3'$ based upon the locally generated random number, and $m_a$ and $H(ID_{edge})$ values as received from the edge device, will not match $M_3$. Likewise, any manipulation on the message ($H(ID_{gw}, M_4, m_b)$) will easily be detected on the edge-node as the value of $M_4'$ will not match $M_4$. Similarly, manipulation of $M_5$ would require knowledge of CSI values for successful decryption and thus is unlikely to occur.  In the continuous phase, the MiTM attacker would need to know the recent $SN_{key}$ value, which depends upon the CSI values, and thus the protocol will be secure against MiTM attacks even during this phase.

\subsection{Mutual Authentication} 
Mutual authentication implies that two communicating parties can authenticate one another. In the mutual authentication phase, the gateway verifies the identity of an edge-node by comparing $M_3'$ and $M_3$ (which is sent by the edge-node). As both of these values are computed by leveraging a random number which is only known to valid devices, a positive match authenticates the edge-device. The edge-node on the other hand matches $M_4'$ and $M_4$ to authenticate the gateway, which again are computed based upon a random number and the key, $E_{key}$. A positive match thus confirms the authenticity of the gateway before agreeing upon a session key $SN_{key}$. During the continuous authentication phase, both edge-node and gateway authenticate each other by leveraging the time-dependent linear or non-linear functions whose exponent values are set securely and dynamically by both devices using the recent generated (fresh) dynamic session key, and thus known only to the two legitimate devices. In addition, they also make use of counter values to validate the sequence of received messages from each other. Therefore, our protocol achieves mutual authentication for both phases.  

\subsection{Cloning Attack} 
Cloning attacks are defined as follows; an attacker creates a replica of the genuine device to launch the attack by masquerading as a valid device. Several methods for creating device replicas are mentioned in the literature \cite{sathyadevan2019protean}. For example, assume that an attacker uses a sophisticated method to craft the replica of a valid device after the mutual authentication phase in our protocol is successfully completed. It appears that this will reveal the $SN_{key}$ value and seed required for the $PRNG$ to the attacker, making it possible for him/her to establish a session with the target device. Prior reported work \cite{chuang2018lightweight, sathyadevan2019protean} also suffers from this type of attack, where keys can be revealed to lead to compromise of the entire protocol. To counter these attacks, existing works assumes that the keys and other minimal information may be stored in a secure location which may not be revealed to the attacker even if the device replica is created. However, our protocol provides an added benefit as the keys are valid only for fixed time duration of \emph{T} seconds. For critical applications, this \emph{T} can be small, making it difficult for an attacker to clone the device within \emph{T} secs, after which keys and random numbers are revoked and regenerated. In addition, even if an attacker succeeds in cloning a device within \emph{T} secs, the copied information will only be valid for a certain time period, thereby requiring the attacker to create a new replica each and every time a valid device changes the $SN_{key}$ and random numbers, which makes it impractical for an attacker.

\subsection{Sybil Attack}
In a Sybil attack, an adversary forges or uses stolen identities of nodes in a peer-to-peer network and tries operating as a legitimate node. The adversary aims to influence the effectiveness of the system by acquiring an unreasonable level of control. Such attacks can affect data integrity and overall reputation of the system. In our proposed  authentication scenario, a Sybil attack can occur if two edge nodes can send the same $ID_{edge}$ packet during the mutual authentication phase. However, as already discussed for impersonation attacks, the protocol is secure against identity forging attacks, as security keys constantly change. In addition, the protocol also does not rely on trust calculations, and authentication of a device is solely based on the identities and keys exchanged between the devices, changing dynamically, and hence is resistant to loss of reputation or trust in the system. Only possible information that a malicious edge node can gather from the mutual authentication phase would be the identity of the gateway, which may not be useful, as security keys do not depend upon device identities. During the continuous authentication phase, no identities are exchanged and authentication depends only on the random numbers and recent security keys exchanged between an edge and the gateway, ascertaining security against such an attack. 

\section{Conclusion \& Future Work} \label{conclusion}
Continuous d2d authentication that uses stable fingerprints and is secure from a variety of protocol attacks is indispensable to successfully instantiate ZTA in critical infrastructures. In this paper, we presented our novel lightweight continuous authentication protocol for device-to-device communication. We also discussed how our protocol is resilient against the most commonly reported attacks in the related literature. In the future, we plan to further expand our informal analysis and also use verification tools to formally prove the security of our proposed protocol.

\section{Acknowledgment}
This work was funded by the Australian Department of Defence through the Defence Science and Technology Group’s Operations Research Network.

\bibliographystyle{IEEEtran}
\bibliography{zta.bib}

\end{document}